\documentstyle [11pt]{article}
\begin{document}
\hfill  ORNL-CCIP-94-08 / RAL-94-056
\vspace{1.0cm}
{
\begin{center}
{\Large\bf THE STATUS OF MOLECULES}
\footnote[1]{Presented at the XXIX Recontres de Moriond, M\'eribel, France
(19-26 March 1994).}\\
\vspace{1cm}
T.Barnes\\
Physics Division and Center for Computationally Intensive Physics\\
Oak Ridge National Laboratory, Oak Ridge, TN 37831\\
and\\
Department of Physics and Astronomy\\
University of Tennessee\\
Knoxville, TN 37996\\

\date{}
\end{center}
}

\begin{abstract}
This report summarizes the experimental and theoretical status of hadronic
molecules, which are weakly-bound states of two or more hadrons. We begin with
a brief history of the subject and discuss a few good candidates, and then
abstract some signatures for molecules which may be of interest in the
classification of possible molecule states. Next we argue that a more general
understanding of $2\to 2$ hadron-hadron scattering amplitudes will be crucial
for molecule searches, and discuss some of our recent work in this area. We
conclude with a discussion of a few more recent molecule candidates (notably
the $f_0(1710)$) which are not well established as molecules but satisfy some
of the expected signatures.
\end{abstract}

\section{Brief History}
\thispagestyle{empty}

In the 1970s it was widely believed that there would be a very rich
spectrum of discrete levels of multiquark resonances. The argument was that
the many known $q\bar q$ and $qqq$ resonances exist because they are
color singlets, so we should expect other color-singlet
sectors of Hilbert space
to possess resonances as well. The ``four-quark"
$q^2\bar q^2$ system was the subject of many detailed studies because
it contains the first color-singlet
multiquark system beyond three quarks, and because
this system could couple to baryon-antibaryon systems through a
single $q\bar q$ annihilation. Partly for this reason $q^2\bar q^2$ states
were referred to as ``baryonia". Although there were many reports of possible
experimental baryonium states, and many detailed spectra were published in
various models, no such states have yet been established. In sectors which
support $q\bar q$ states the spectrum is already very complicated, so the
issue of multiquark states remains somewhat obscure.
However when one specializes to ``smoking gun"
systems such as the exotic
I=2 channel, which is predicted to support a light $0^{++}$
$q^2\bar q^2$ level (at about 1.2 GeV in the MIT bag model) but cannot have a
$q\bar q$ state, there is no resonance in evidence \cite{Hoogland}.

The problems with the various theoretical models that led to
erroneous predictions
of discrete multiquark levels
have been discussed by Isgur
\cite{NI}.
The novel feature of multiquark systems
which the models missed is that, unlike $q\bar q$ and $qqq$,
they need not exist as
single color-singlet hadronic clusters;
a $q^2\bar q^2$ system in general has some projection onto
two color-singlet $q\bar q$ mesons,
and continuous deformation into
two separate mesons appears to be energetically favored
in most cases. This rearrangement into color singlets is called ``fall-apart"
\cite{fall},
and apparently excludes most single-hadron $q^2\bar q^2$ clusters as
resonances.
Fall-apart would not be possible if the cluster had a
mass lower
than the threshold of the
two-hadron system it can rearrange into, which is why the
question of the existence of multiquark clusters such as the H $u^2d^2s^2$
system is so interesting. The bag model predicts this state
81 MeV below $\Lambda\Lambda$ threshold \cite{fall},
but this prediction should be treated with caution because the bag model has
previously given a misleading picture of multiquark states.
The tentative evidence
for dilambda hypernuclei \cite{LL}
(if confirmed) makes the existence of
an H six-quark resonance
well below $\Lambda\Lambda$ threshold appear very unlikely.
Whether
single multiquark clusters
exist as resonances under any conditions is a detailed dynamical question,
which should be investigated using models that allow the system itself
freedom to choose between a single cluster or separate color singlets.
At present it appears that single $q^2\bar q^2$ hadronic clusters
may only exist as resonances in heavy-light systems such as $c^2\bar q^2$
\cite{Richard}.

It was the lack of sufficient
freedom in the wavefunctions that led to the spurious prediction of
many discrete
baryonium levels; the models {\it assumed} that such states existed, and then
gave predictions for the spectrum of these discrete levels. The first detailed
study that allowed the $q^2\bar q^2$ system to choose between clusters and
separate mesons as
ground states
was the variational study of the $0^{++}$ sector
by Weinstein and Isgur \cite{WI}, which found that
continuous deformation of a cluster into separate
$q\bar q$ mesons was usually preferred energetically.
The two exceptions found by Weinstein and Isgur will be discussed
below.

\section{A Few Good Candidates}

\noindent
2.1) Nuclei
\vskip 0.3cm

Lest one form the impression that hadronic molecules are controversial,
note that the $\sim 10^{5}$ known nuclear levels are all
hadronic molecules. Of course the term is usually applied to hadron pairs;
even if we specialize to this restricted case,
the deuteron can be cited as a noncontroversial example of a dominantly S-wave
hadronic molecule.
Its almost-bound I=1, S=0 partner is an example
of another phenomenon which may appear elsewhere in the spectrum, a
molecular
resonance above threshold
which is due to a strongly attractive final-state interaction.
The existence of so many nuclear species is especially notable since the
nucleon-nucleon system is rather unfavorable for the formation of bound states,
due to the strong short-distance repulsive core. This
suggests that many other families of bound hadrons may exist, although they may
not be experimentally accessible except in special cases.

\vskip 0.5cm
\noindent
2.2) $\Lambda(1405)$
\vskip 0.3cm

In meson-baryon sectors, the ${1\over 2}^-$
$\Lambda(1405)$ has long been considered a candidate
$\bar K N$ bound state \cite{DT},
since it is just below the $\bar KN$ threshold,
has S-wave $\bar KN$ quantum numbers, and is nearly 150 MeV
below the mass expected for
the lightest ${1\over 2}^-$
$\Lambda^*$ baryon in the quark model \cite{CI}.
(The decay amplitudes however are consistent with a $uds$
assignment \cite{CI}.)
Theoretical study of this channel has been incomplete, however,
because of the complications of open channels and $q\bar q$ annihilation;
a model which includes mixing in the full $N\bar K$-$\Sigma\pi$-$uds$ system
is required for a complete study of this state.
The
radiative partial widths to $\gamma\Lambda$ and $\gamma\Sigma^0$ compared to
quark model predictions for a $uds$ baryon may allow a convincing test
of the molecule assignment. A wide range of
theoretical numbers has been reported for these
radiative widths \cite{Lowe}, so one should be careful to use techniques
which give reliable results for well-established $qqq$ quark model
states.

\vskip 0.5cm
\noindent
2.3) $f_0(975)$ and $a_0(980)$
\vskip 0.3cm

Weinstein and Isgur \cite{WI} found an exception to the fall-apart phenomenon
in the scalar sector, with parameters corresponding to the $qs\bar q\bar s$
system.
Here
weakly-bound deuteronlike states of kaon and antikaon were found to be the
ground states of the four-quark system; Weinstein and Isgur refer to these as
``K$\bar{\rm K}$ molecules". The scalars
$f_0(975)$ and $a_0(980)$
were obvious
candidates for these states, having masses just below $K\bar K$
threshold and strong couplings to strange final states. Subsequently the
$\gamma\gamma$ couplings of the
$f_0(975)$ and $a_0(980)$
were found to be anomalously small
relative to expectations for light $^3P_0$ $q\bar q$ states ($q=u,d$), as
discussed in Refs.
(\cite{MRP,ggKK}).
The status of the $K\bar K$ molecule assignment and the many points of evidence
in its favor have been discussed recently by Weinstein and Isgur
\cite{WInew,Weinstein}.

Although Morgan and Pennington have argued against a molecule interpretation
of the $f_0(975)$ \cite{MP}, their criticism applies to a $K\bar K$ potential
model in which the $f_0(975)$ is a single pole in the scattering amplitude.
The more recent work of Weinstein and Isgur
\cite{WInew,Weinstein} incorporates couplings to
open meson-meson channels and heavier $^3P_0$ $q\bar q$ states, and although
the $f_0$ and $a_0$ states remain dominantly $K\bar K$, these modifications
may answer the objections of
Morgan and Pennington.  Pennington suggests that the term
``deuteronlike" may be a misnomer, if couplings to other states than $K\bar K$
play an important r\^ole in these states \cite{MRP}. Thus it appears that the
important question regarding the $f_0$ and $a_0$ may be one of detail,
specifically
how large the subdominant non-$K\bar K$ components are in these states and how
they can be observed experimentally.

Finally, Gribov and collaborators \cite{Gribov}
have discussed the possibility that the $f_0(975)$
and $a_0(980)$ might be novel $q\bar q$ states constructed of negative-energy
Dirac levels, which they expect might have small length scales of $\sim 0.2$ fm
and mass scales of $\sim 1$ GeV. They note that this suggestion can also
be tested by an accurate determination of the $\gamma\gamma$ partial widths
of these states.

\section{Signatures for Molecules}

Leaving aside questions of detailed dynamics, there are several obvious
signatures
for hadron-pair molecules that may be abstracted from these candidates.
These signatures are:

\vskip 0.5cm
\noindent
1) {\bf $J^{PC}$ and flavor quantum numbers of an L=0 hadron pair.}

\vskip 0.5cm
\noindent
The residual ``nuclear" strong forces that bind molecules are of such short
range that L$>$0 molecules appear unlikely in light hadronic systems.
Actually there is a possible exception, the $\psi(4040)$, which couples so
strongly to $D^*\bar D^*$ that it was suggested as a P-wave $D^*\bar D^*$
molecule (note $E_B\approx 0$) some time ago \cite{Novikov}.
This exception may be possible because heavier mesons bind more
easily, and the light quarks insure relatively strong interactions between
them.
Should this state actually be a P-wave charmed-meson molecule, a spectrum
of more deeply-bound S-wave charm molecules is anticipated \cite{Georg}.

\vskip 0.5cm
\noindent
2) {\bf A binding energy of at most about $50-100$ MeV.}
\vskip 0.5cm
\noindent
{}From the uncertainty principle; a minimum separation
of $\approx 1$ fm is required for hadrons
to maintain separate identities, which gives $E_B\sim 1/(2\mu R^2) \sim 1/(1
\;\hbox{GeV}) \cdot (1 \;\hbox{fm})^2\approx 50 \;\hbox{MeV}$, and of course
a factor-of-two uncertainty is plausible in this simple estimate.
For comparison,
Weinstein and Isgur find an rms $K\bar K$ separation
of about 1.7 fm in their model
of the $f_0(975)$ and $a_0(980)$, which have $E_B\approx 10-20$ MeV. Note
also that an attractive interaction may lead to a
final-state enhancement in
S-wave just {\it above} threshold, which may or may not be resonant; the
I=1, S=0 partner of the deuteron is an example of such a resonance,
and the
$f_1(1420)$ may be an example of a nonresonant final-state enhancement.

\vskip 0.5cm
\noindent
3) {\bf Strong couplings to constituent channels.}
\vskip 0.5cm
\noindent
As an example, the
anomalously large coupling of the $f_0(975)$ to $K\bar K$, as indicated by
$B(K\bar K) / B(\pi\pi) \approx 1/4$ despite the near absence of $K\bar K$
phase space, is an important clue that it is not a nonstrange $q\bar q$ state.

\eject
\noindent
4) {\bf
Anomalous EM couplings relative to expectations for conventional quark model
states.}
\vskip 0.5cm
\noindent
The $f_0(975)$ for example has a tiny $\gamma\gamma$ partial width of perhaps
0.2 Kev to 0.6 Kev (depending on the analysis) \cite{MRP}.
This small $\gamma\gamma$ width is expected for a $K\bar K$ molecule
\cite{ggKK}, but
for a nonstrange $^3P_0$ $q\bar q$ state the quark model predicts
about
3 Kev
\cite{gammas}, as has recently
been found for the broad $f_0(\approx 1300)$ \cite{CBepsilon}. Close, Isgur and
Kumano
\cite{CIK} suggest a related test
for the $f_0(975)$ and $a_0(980)$
involving
the radiative decays $\phi\to\gamma (f_0,a_0)$,
which may be possible at DA$\Phi$NE
and CEBAF.
\vskip 0.5cm

\section{Back to Basics: $2\to 2$ Scattering Amplitudes}

Low-energy
$2\to 2$ hadron scattering
is interesting in itself as a nontrivial aspect of QCD,
and if we can reach an
understanding of the important scattering mechanisms
in terms of quarks and gluons,
we should be able to predict which channels
experience strong attractive forces
and hence may support molecular bound states.

Hadron-hadron scattering amplitudes at low energies
are generally thought to involve the quark-gluon interaction nonperturbatively,
so although calculations of meson-meson and baryon-baryon interactions
at the quark-gluon level have been rather successful, they have typically
used complicated nonperturbative methods such as resonating group
or variational techniques. Extension of this work to channels such as
vector-vector has been slow largely because of the difficulty of applying
these
methods, although some variational and Monte Carlo results have been reported
for special cases, including an extension of the
Weinstein-Isgur work to I=2 $\rho\rho$ \cite{wigg}.

Recently our collaboration has found evidence that the Ps-Ps scattering
amplitudes found by Weinstein and Isgur
(in channels without $q\bar q$ annihilation) are actually dominated by
perturbative diagrams, although ``higher-twist" contributions in the
form of external $q\bar q$ wavefunctions attached to the diagrams
are an essential,
nonperturbative
aspect of the scattering amplitudes \cite{BS,Swan}.
We initially studied I=2 $\pi\pi$ \cite{BS} and
I=3/2 $K\pi$ \cite{BSW} and found that OGE followed by constituent
interchange dominates these scattering amplitudes, and leads to
results which are numerically very similar to the Ps-Ps potentials
found variationally by Weinstein and Isgur.
Recent lattice QCD results for the I=2 $\pi\pi$ scattering length support
our conclusion regarding the dominance of these diagrams \cite{Sharpe}.
We refer to these perturbative
diagrams with external wavefunctions attached
as ``quark Born diagrams".
With SHO quark model wavefunctions
these lead to overlap integrals that can often be evaluated in
closed form, and the
results for $\pi\pi$ and $K\pi$
S-wave phase shifts are in excellent agreement with experiment over the
entire range of energies studied given standard quark model parameters.
We have similarly found good agreement
in the I=0 and I=1
KN system \cite{BSKN}
({\it albeit} with some problems at higher energies
which may be due to the assumption of single-Gaussian
nucleon wavefunctions).
In our study of NN, N$\Delta$ and $\Delta\Delta$
\cite{BCKS}
we found the strongest diagonal attraction in the
$I=0,S=1$ $\Delta\Delta$ channel,
in agreement with the
variational work of Maltman
\cite{Maltman}.
Very recently we have
studied the N$_s=2$ baryon-baryon channel \cite{web}, and we agree with
Oka, Shimizu and Yazaki and Straub {\it et al.}
\cite{LLthy} that quark model forces lead to a
repulsive $\Lambda\Lambda$ core interaction;
of the six
N$_s=2$ octet-octet channels we find that only I=0,S=0
$\Sigma\Sigma$ has an attractive core.
It is reassuring that exactly the same conclusion
was reached by
Oka {\it et al.}
using nonperturbative resonating group methods.

Given this reasonably successful description
of hadron-hadron scattering at low energies, can we
proceed to study all experimentally accessible channels and see in which we
expect hadronic molecules to form? Unfortunately this is not yet possible.
The difficulty is that
$q\bar q$ annihilation
appears to be the dominant effect
when allowed (as in I=0,1 $\pi\pi$ and
I=1/2 $K\pi$),
so a realistic
description of scattering amplitudes in these channels
requires accurate modelling of the couplings of different sectors
of Hilbert space (as in $\pi\pi\to f_0(q\bar q)\to\pi\pi$).
On closer examination of the
Weinstein-Isgur results \cite{Weinstein} it now appears that
both level repulsion against higher-mass
s-channel $q\bar q$ resonances and nonresonant scattering are needed to
bind both the I=0 and I=1 $K\bar K$ systems. Without
level repulsion against the
$q\bar q$ $^3P_0$ states at $\approx 1.3$ GeV only the I=0 molecule binds.
The existence of molecules in some channels may be due entirely
to level repulsion against
a more massive quark model state, and the hadronic couplings of
most quark-model states
are not well enough established to model this level repulsion accurately.
Remarkably,
the $q\bar q$ pair production process is still rather poorly understood
(for a recent study see \cite{GS}),
and is usually treated using phenomenological models such as $^3P_0$ that
have no clear relation to QCD.
Accurate modelling of hadronic forces in channels with
annihilation, and hence reliable predictions of molecules,
must await a better understanding of the $q\bar q$
annihilation mechanism.

\section{A Few More Candidates}

There are many resonances above 1.4 GeV
which may prove to be
molecular states.
Here I will discuss
three such
I=0 resonances which do not appear to have a natural assignment
as $q\bar q$ mesons, and which merit consideration as
molecule candidates. These are
the ``$\theta(1720)$" (now known as the $f_0(1710)$), the
$AX(1515)$ (originally reported by ASTERIX
$P\bar P\to \pi^+\pi^-\pi^0$ \cite{May}, now believed
to be dominantly J=0 and called the $f_0(1520)$ \cite{CBar}),
and the $f_1(1420)$. For completeness I conclude with a
reminder of other sectors
of Hilbert space which may possess molecular resonances.

\vskip 0.5cm
\noindent
5.1) $f_0(1710)$
\vskip 0.25cm

Early
references often considered this a glueball candidate, since it was
discovered in a $\psi$ radiative decay \cite{Edwards} and has no
obvious assignment in the $q\bar q$ spectrum. It was also discussed as a
possible $qs\bar q\bar s$ multiquark state (a single cluster rather than a
molecule), although the presence of a fall-apart coupling to $K\bar K$ makes
this appear untenable. The $f_0(1710)$ appears unlikely to be a radially
excited nonstrange $^3P_0$ $q\bar q$ because it has a mass 110 MeV below the
Godfrey-Isgur prediction \cite{GI} and it has a very weak coupling to $\pi\pi$
final states, which makes both nonstrange $q\bar q$ and
glueball assignments rather implausible
(assuming naive flavor-singlet glueball couplings).

We consider the $f_0(1710)$ a strong vector-molecule candidate.
There have been several suggestions in the literature regarding
molecule assignment for this state, which differ primarily in the
proposed binding mechanism;
this leads to observable differences in the predicted
decay modes.

T\"ornqvist \cite{Torn} suggests a $K^*\bar K^*$ assignment for this state. If
the $K^*$s decay as free hadrons this leads one to expect a large partial width
of $\Gamma(f_0(1710)\to K^*\bar K^*) \approx 2\, \Gamma(K^*)\approx 100$ MeV,
making it the dominant decay mode, since the 1992 PDG gives the $f_0(1710)$ a
total width of $146\pm 12$ MeV. This prediction appears to disagree with the
branching fractions to other final states estimated by the PDG. Ericson and
Karl \cite{EK} have studied the one-pion-exchange
binding mechanism
proposed by T\"ornqvist,
and conclude that it may just provide sufficient
attraction to bind $K^*{\bar K}^*$.
They also conclude that this mechanism would predict several other
vector-vector and meson-baryon molecules, such as $B^*B^*$
and $N\Sigma$.

Dooley, Swanson and Barnes \cite{DSB}
assumed that a different mechanism was responsible for the dominant
vector-vector interactions, specifically a OGE constituent interchange model
which they had previously applied to several other channels (see previous
section). This model predicts a very strong coupling between $K^*\bar K^*$
and $\omega\phi$ channels, found by
Swanson \cite{Swan} and represented using off-diagonal hadron-hadron
potentials. Dooley {\it et al.} used these
potentials
in a multichannel generalization of the Weinstein-Isgur
work, and found weakly bound molecules in several vector-vector systems.
In the
I=0 $qs\bar q\bar s$ sector they found a linear combination of two-meson basis
states as a $0^{++}$ ground state,
\begin{equation}
|\Psi_0(qs\bar q\bar s\ 0^{++})\rangle = {1\over \sqrt{2}}
\bigg( |K^*\bar K^*\rangle + |\omega\phi\rangle \bigg)\ ,
\end{equation}
and a $2^{++}$ excited state somewhat higher in mass.
This linear combination predicts $K\bar K$ and
$\eta\eta$ branching fractions close to experimental values \cite{Dooley} and
also gives reasonable results for the flavor-tagging $\psi\to VX$ hadronic
decays, which suggest that the $f_0(1710)$ couples as if it were a
strange-nonstrange mixed-flavor state.
(A similar result is known for the $f_0(975)$ in these decays.)
A $K^*\bar K^*$ decay mode is expected
as in the pure $K^*\bar K^*$-molecule picture, but with a smaller branching
fraction of about 35\%.
The branching fraction of a $(K^*\bar K^* +
\omega\phi)/\sqrt{2}$ molecule to $\pi\pi$
is more problematical; Dooley has recently found it to be about
5\% \cite{Dooley}, consistent with experiment.
A very characteristic electromagnetic decay mode in this assignment is
$f_0(1710)\to\phi\pi^0\gamma$, with a
branching fraction of 0.3\%, due to constituent-$\omega$
radiative decay. It should also have an anomalously small $\gamma\gamma$
coupling relative to a nonstrange I=0 $q\bar q$ state such as the $f_2(1274)$
\cite{thetagams}.

A search for a $K^*\bar K^*$ mode is clearly the most important test of the
vector-molecule models of the $f_0(1710)$, since both models expect a large
$K^*\bar K^*$ branching fraction. Of course one $K\pi$ combination will be
skewed downwards from the free $K^*$ mass by slightly more than $E_B$, assuming
that this acts as a sequential decay, $f_0(1710)\to K^* (\bar K \pi)$ followed
by decay of the free recoiling $K^*$.

\vskip 0.5cm
\noindent
5.2) $f_0(1520)$
\vskip 0.25cm

This state, like the $f_0(1710)$, has a complicated history in which it was
originally attributed to a $2^{++}$ resonance \cite{May,AX}, with
subsequent analysis changing it to $0^{++}$ with a small additional $2^{++}$
amplitude \cite{CBar}. The recent Crystal Barrel study
of the reactions
$P\bar P \to \pi^o\pi^o\pi^o$
and
$P\bar P \to \eta\eta\pi^o$ finds two broad scalar states, an
$f_0(1365)$ with a mass and width of
$M=1365{+20\atop -55}$ MeV and
$\Gamma=268\pm 70$ MeV
(which appears consistent with expectations for a $^3P_0$ $q\bar q$ state)
and the
$f_0(1520)$, with a mass and width of
$M=1520\pm 25$ MeV and
$\Gamma=148{+20\atop -25}$ MeV.
The mass of the $f_0(1520)$
rules out any quarkonium
assignments except $s\bar s$, but
its couplings are inconsistent with $s\bar s$;
it has been reported in
$\pi\pi$, $\eta\eta$ \cite{CBar},
and $\eta\eta'$ final states \cite{Hack}, with
amplitudes approximately consistent with flavor-singlet couplings.

The reported couplings of the $f_0(1520)$ make it
a plausible glueball candidate,
as does its mass; recent lattice gauge theory expectations for the
lightest glueball are that it should be a scalar with a mass of about
1.5 GeV \cite{Michael}.
Of course this prediction uses the quenched approximation,
but this
leads to quite reasonable results for the spectrum of
conventional light $q\bar q$ mesons. A crucial test of the glueball
assignment will be a measurement of the branching fraction to $K\bar K$;
if this is also consistent with a flavor singlet, then a strong case for
the identification of the $f_0(1520)$ with the light scalar glueball can
be made.

If the $f_0(1520)$ is found to couple only weakly to $K\bar K$, another
possibility is that it is a vector meson molecule.
Since the $\rho\rho$ and $\omega\omega$ thresholds are not far above this state
and it
has S-wave $\rho\rho$ and $\omega\omega$ quantum
numbers, it is an obvious candidate for a vector-vector molecule.
A possible $\rho\rho$ assignment has been suggested by
both Kalashnikova \cite{kalash} and T\"ornqvist \cite{Torn}.
A pure $\rho\rho$ bound state however appears inconsistent with the new
Crystal Barrel width, since a weakly bound $\rho\rho$ state would have a width
due to constituent decay of
\begin{equation}
\Gamma(\rho\rho) \approx 2\, \Gamma(\rho) \approx 300 \;\hbox{MeV} \ .
\end{equation}

Another possibility is that strong mixing between the nearly degenerate
$|\rho\rho\rangle$ and $|\omega\omega\rangle$ basis states has led to a
coherent superposition close to
\begin{equation}
|VV\rangle = {1\over \sqrt{2}}
{\bigg( |\rho\rho\rangle + |\omega\omega\rangle\bigg) } \ ;
\end{equation}
with this linear combination one would expect a strong width from constituent
decays alone of
\begin{equation}
\Gamma(VV) \approx \Gamma(\rho) +\Gamma(\omega)
\approx 160 \;\hbox{MeV} \ ,
\end{equation}
consistent with the reported total width of the $f_0(1520)$.
This
$(\rho\rho + \omega\omega)/\sqrt{2}$
model also makes several other characteristic predictions,
such as the dominance of $\rho\rho$ decays,
$ B(\omega\omega)/B(\rho\rho) \approx
{1/20}$ and
$\Gamma (f_0(1520)\to \omega\pi^0\gamma)
\approx 0.8$ MeV
(both from consideration of constituent decays).

The possibility of nonstrange vector-vector molecules may have
independent support
from the
Crystal Barrel collaboration
\cite{rhorho},
who report evidence for a
$\rho\rho$ enhancement with a mass and width of
$M=1374\pm 38$ MeV and $\Gamma=375\pm 61$ MeV. These values are
inconsistent with the
$f_0(1520)$ alone, and might be due to the $f_0(1365)$, to a combination of
the $f_0(1365)$ and $f_0(1520)$, or even to a third broad $f_0$.

\vskip 0.5cm
\noindent
5.3) $f_1(1420)$
\vskip 0.25cm

The final unusual meson state we consider
is the $f_1(1420)$,
which is a candidate for
a nonresonant threshold
enhancement
($K^*\bar K + h.c.$)
rather than a molecular bound state.
This possibility was suggested by
Caldwell \cite{Caldwell}, and satisfies the criteria of lying just above
the $K^*\bar K$
threshold
and having quantum numbers allowed for that pair in S-wave.
The apparent width of the enhancement should not be narrower than the intrinsic
width of the $K^*$, and indeed the PDG values are similar,
$\Gamma(f_1(1420))=56\pm 3$ MeV and $\Gamma(K^*)=50$ MeV. Longacre \cite{Long}
found that a model with an S-wave nonresonant $(K^*\bar K+h.c.)$ enhancement
gives a good description of this state, and Isgur, Swanson and Weinstein
\cite{ISW} also favor this possibility. The (off-shell) $\gamma\gamma^*$
couplings of the $f_1(1420)$ relative to expectations for a $1^{++}$ $s\bar s$
state may provide a test of the hadron-pair model.

\vskip 0.5cm
\noindent
5.4) $Z^*s$ and dibaryons
\vskip 0.25cm

We conclude this section with a reminder that there may be hadronic
molecules in other sectors of Hilbert space, which have received little recent
attention because they do not correspond to $q\bar q$ or $qqq$ flavor states
in the quark model or because of the lack of appropriate experimental
facilities.

One especially interesting system, which should soon be accessible to
experiments at DA$\Phi$NE and perhaps CEBAF, is the kaon-nucleon system.
Possible resonances in the $q^4\bar s$ sector are known as $Z^*$s, and
although there have been indications of such states for many years
in elastic KN scattering \cite{VPIKN},
the lack of clear evidence for multiquark states and the uncertainties of
partial wave analyses in the $KN$ system have left the possibility of such
states a controversial question.
In our recent theoretical study of $KN$, $K^*N$, $K\Delta$ and $K^*\Delta$
systems \cite{BSKN}
we found that several of these channels,
notably the minimum-total-spin, minimum-total-isospin ones,
have strongly attractive interactions. The experimental reports of
$Z^*$ resonances may represent observations of final state enhancements
or even of meson-baryon bound states just below threshold. Clarification
of this issue will require accurate partial wave analyses of $KN$ scattering,
and data on the inelastic channels $KN\to K^*N, K\Delta $ and $K^*\Delta$
would also be very valuable for the study of possible $Z^*$ states.
Here the theoretical calculations should be more reliable since these
reactions are
annihilation-free at valence quark level.

Finally, there are controversial reports of
resonances in partial wave analyses of
elastic
NN scattering \cite{VPINN}, and these
``dibaryon" resonances
may also include S-wave
baryon-baryon molecule states. This system too is
relatively straightforward
theoretically because it is annihilation-free, and
a few
$\Delta\Delta$ channels (notably I=0,S=1, I=1,S=0, I=0,S=3 and
I=3,S=0) have
been cited by theorists as the most likely for the formation of nonstrange
baryon-baryon resonances \cite{BCKS,Maltman}.

\section{Summary and Conclusions}

The hadronic spectrum exhibits many
quasinuclear hadron bound states,
which have become known as ``molecules". In this talk we discussed the status
of several of these candidate hadronic molecules, including nuclei
(which are nucleon molecules), the $\Lambda(1405)$,
the mesons $f_0(975), a_0(980), f_0(1710), f_0(1520)$ and $f_1(1420)$,
$Z^*$s and dibaryons.

The study of molecules is a subtopic of the problem of determining
$2\to 2$ hadron-hadron scattering amplitudes near threshold. Although this
is widely held to be a nonperturbative problem,
our collaboration has found that a simple class of perturbative diagrams
(with external quark wavefunctions attached)
dominates low-energy scattering in annihilation free channels.
We refer to these diagrams as
quark Born diagrams; their study has led us to predictions of several
channels which may support hadronic molecules. One such state
is the $f_0(1710)$,
which we believe may be a vector-vector $(K^*\bar K^*+\omega\phi)/\sqrt{2}$
molecule;
this assignment leads to detailed predictions of couplings and decay
modes.

The study of molecules would be greatly assisted by a better understanding
of hadron scattering mechanisms at the quark and gluon level.
For this reason we particularly advocate future
studies of scattering amplitudes in channels such as KN,
in which one can study nonresonant two-body scattering in the absence of
complications due to valence $q\bar q$ annihilation.

\section{Acknowledgements}

It is a pleasure to acknowledge the efforts of the organisers of this meeting,
in particular Tran Thanh Van and Lucien Montanet. The
contributions of my collaborators S.Capstick, F.E.Close,
K.Dooley, G.Grondin, M.D.Kovarik, Z.P.Li, E.S.Swanson and
J.Weinstein
are also gratefully acknowledged.
This research was sponsored in
part by the United States Department of Energy under contract
DE-AC05-840R21400, managed by
Martin Marietta Energy Systems, Inc, and by the United Kingdom Science Research
Council through a Visiting Scientist grant at Rutherford Appleton Laboratory.

\end{document}